\newif\ifmulticol	\multicoltrue
\newif\ifshowgit	\showgittrue		
\newif\ifgitlocal	\gitlocaltrue		
\newif\ifbiblatex	\biblatexfalse		
\newif\ifbibnum		\bibnumfalse 		
\newif\iflineno		\linenofalse
\newif\iftoc		\tocfalse

\newif\iflucida		\lucidafalse
\newif\ifcm			\cmtrue
\newif\iflibertine	\libertinefalse
\newif\ifcharter	\charterfalse

\newcommand*{\secnumdepth}{0}			
\newcommand*{\tocdepth}{2}				
\newcommand*{\reftitle}{References}		
\newcommand*{\mytitleskip}{-0.2in}		

\multicoltrue\showgittrue\gitlocaltrue\biblatexfalse\bibnumtrue

\newcommand*{\mydocfontsize}{\ifcharter11pt\else\iflibertine11pt\else10pt\fi\fi}
\renewcommand*{\mydocfontsize}{\ifcharter11pt\else\iflibertine11pt\else11pt\fi\fi}
\newcommand*{\setcol}{\ifmulticol twocolumn\else onecolumn\fi}

\documentclass[\mydocfontsize,\setcol]{article}

\newcommand*{\pdfauthor}{Steven A.~Frank}
\newcommand*{\pdftitle}{The generalized Price equation: forces that change population statistics}

\input pricegen.sty

\newcommand{\bdq}{\GD\mathbf{q}}
\newcommand{\bq}{\mathbf{q}}

\newcommand{\bdz}{\GD\mathbf{z}}
\newcommand{\bz}{\mathbf{z}}

\newcommand{\zbar}{\bar{z}}

\newcommand*{\GD}{\Delta}

\DeclarePairedDelimiter\abs{\lvert}{\rvert}
\DeclarePairedDelimiter\norm{\lVert}{\rVert}
\DeclarePairedDelimiter\angb{\langle}{\rangle}
\DeclarePairedDelimiter\lrb{\lbrack}{\rbrack}
\DeclarePairedDelimiter\lr{\lparen}{\rparen}

\makeatletter
\let\oldabs\abs \def\abs{\@ifstar{\oldabs}{\oldabs*}}
\let\oldnorm\norm \def\norm{\@ifstar{\oldnorm}{\oldnorm*}}
\let\oldangb\angb \def\angb{\@ifstar{\oldangb}{\oldangb*}}
\let\oldlrb\lrb \def\lrb{\@ifstar{\oldlrb}{\oldlrb*}}
\let\oldlr\lr \def\lr{\@ifstar{\oldlr}{\oldlr*}}
\makeatother

\newcommand*{\Eq}[1]{eqn~\ref{eq:#1}}

\newcount\BoxNum \BoxNum 1\relax
\makeatletter
\newcommand*{\boxlabel}[1]{%
  \protected@write \@auxout {}{\string \newlabel {box:#1}{{\the\BoxNum}}{}}%
  \advance\BoxNum 1\relax}
\makeatother

\newcommand*{\mytitle}{\pdftitle}

\newcommand*{\myauthor}{Steven A.\ Frank}
\newcommand*{\myauthorb}{William Godsoe}
\newcommand*{\myaddress}{Department of Ecology and Evolutionary Biology, University of California, Irvine, CA 92697--2525, USA}
\newcommand*{\myaddressb}{Bio-Protection Research Centre, Lincoln University, P.O.\ Box 85084, Lincoln 7647, New Zealand}

\setabstract{-0.07}{\iftoc-2.0\else1.22\fi}{%
The Price equation partitions the change in the expected value of a population measure. The first component describes the partial change caused by altered frequencies. The second component describes the partial change caused by altered measurements. In biology, frequency changes often associate with the direct effect of natural selection. Measure changes reflect processes during transmission that alter trait values. More broadly, the two components describe the direct forces that change population composition and the altered frame of reference that changes measured values. The classic Price equation is limited to population statistics that can expressed as the expected value of a measure. Many statistics cannot be expressed as expected values, such as the harmonic mean and the family of rescaled diversity measures. We generalize the Price equation to any population statistic that can be expressed as a function of frequencies and measurements. We obtain the generalized partition between the direct forces that cause frequency change and the altered frame of reference that changes measurements.
}

\begin{document}

\mymaketitle

\iftoc\mytoc{-24pt}{\newpage}\fi

\section{Introduction}

The Price equation has been used in many disciplines to study the forces that cause change in populations \autocite{frank18the-price}. In evolutionary biology, the equation separates the changes in traits caused by natural selection from the changes in traits that arise during transmission \autocite{gardner20prices}. The abstract expression of change often highlights general principles. For example, the Price equation shows that certain patterns of correlation between individuals provide sufficient statistics for the evolutionary forces that influence altruistic behaviors, the basis for kin selection theory \autocite{hamilton70selfish,frank98foundations}. 

The Price equation has also been used in ecology to partition change in diversity indices into direct and indirect processes \autocite{godsoe19transmission}. Economists have used the equation to model the evolution of firms \autocite{metcalfe98evolutionary,andersen04population}. In physics and other disciplines, the Price equation neatly separates change caused by directly acting forces from change that arises indirectly from an altered frame of reference \autocite{frank18the-price}. The direct forces and the frame of reference are abstractions of the biologists' components of natural selection and transmission.

A recent collection of articles celebrates the 50th anniversary of the Price equation and summarizes applications across a wide variety of disciplines \autocite{lehtonen20fifty}.

The classic Price equation describes change in any population statistic that can be written in the form of an expected value. Many population statistics cannot be written in that form \autocite{bullen03handbook}. 

This article generalizes the Price equation to handle most other forms of population statistics. We end up with the same generic partitioning of population change into two components: the forces that alter the frequencies of the entities that make up the population holding constant the measurements on each entity and the forces that alter the measurements on each entity holding constant the frequencies \autocite{frank12naturalb}.

\section{The Classic Price Equation}

Consider a population statistic that can be written as the expected value of some measurement
\begin{equation*}
  \zbar=\sum_i q_iz_i=\bq\cdot\bz.
\end{equation*}
The index $i$ denotes subsets of the population. The frequencies $q_i$ describe the weighting of each subset. The measurements $z_i$ are any value that may be associated with subsets. The statistic $\zbar$ is the expected value of the measurement. 

Vector notation provides a more compact expression. The vectors $\bq$ and $\bz$ run over the indices $i=1,2,\dots,n$. The dot product, $\bq\cdot\bz$, yields the summation that defines the expected value.

For any population statistic that can be written as an expected value, the classic form of the Price equation partitions the total change in the statistic into two components. If we define the statistic in the altered population as $\zbar'=\bq'\cdot\bz'$, then
\begin{align*}
  \GD\zbar&=\bq'\cdot\bz'-\bq\cdot\bz\\
          &=\bq'\cdot\bz'-\bq'\cdot\bz+\bq'\cdot\bz-\bq\cdot\bz\\
          &=\bq'\cdot\bdz+\bdq\cdot\bz.
\end{align*}
It is convenient to switch the order of the terms in the last line
\begin{equation}\label{eq:canonical}
  \GD\zbar=\bdq\cdot\bz+\bq'\cdot\bdz.
\end{equation}
This expression provides the most basic form of the classic Price equation. Frank\autocite{frank12naturalb} relates this vector form to other notations, such as the covariance and expectation used by Price\autocite{price70selection,price72extension}. 

Equation \ref{eq:canonical} follows the strictly defined set mapping scheme for the relations between entities in the two populations given in Frank\autocite{frank12naturalb}. For example, $q_i'$ is the frequency of entities in the altered population derived from entities in the original population with index $i$. Our extension in the following section also adheres to that strictly defined set mapping. See Frank\autocite{frank12naturalb} for further details.

The Price equation is simply the chain rule for differences applied to a population statistic. The first term on the right-hand side of \Eq{canonical} describes the change in frequencies when holding the measurements constant. The second term describes the change in the measurements when holding the frequencies constant. In the second term, the constant frequencies come from the altered population, because the consequences of the changed frame of reference depend on the frequencies in the altered population. We can express these partial changes with the notation
\begin{equation}\label{eq:partialz}
  \GD\zbar=\GD_q\zbar+\GD_z\zbar,
\end{equation}
in which $\GD_q$ is the partial change with respect to $q$, and $\GD_z$ is the partial change with respect to $z$ when evaluated in the altered population. See Frank\autocite{frank18the-price} for the abstract mathematical properties of the classic Price equation and applications of that abstract interpretation. 

In biology, the partial frequency changes, $\GD_q$, often associate with natural selection, which directly changes frequencies in proportion to Fisher's average excess in fitness. The partial measure changes, $\GD_z$, often associate with changes in trait values during transmission. Thus, the Price equation provides a general separation between selection and transmission.

\section{The Generalized Price Equation}

Many population statistics cannot be expressed as the expected value of a particular measurement. Consider any population statistic that can be written as $f(\bq,\bz)$, where $f$ is a function that maps the population frequencies and measurements to some value. For example,
\begin{equation}\label{eq:inverseMean}
  f(\bq,\bz)=\frac{1}{\sum_iq_iz_i}.
\end{equation}
Many population statistics depend on the frequencies and values across population subsets \autocite{bullen03handbook}. So we should be able to write most population statistics as some function, $f$. The change in $f$ is
\begin{equation*}
  \GD f = f(\bq',\bz')-f(\bq,\bz).
\end{equation*}
Applying the same approach as in the derivation of the classic Price equation, we can add $f(\bq',\bz)-f(\bq',\bz)=0$ to the expression above and then re-arrange to obtain
\begin{equation}\label{eq:general}
  \GD f = \lrb{f(\bq',\bz)-f(\bq,\bz)}+\lrb{f(\bq',\bz')-f(\bq',\bz)},
\end{equation}
which may be expressed in the partial difference notation as
\begin{equation*}
  \GD f = \GD_q f+\GD_z f.
\end{equation*}

\section{Example}

Using the example function in \Eq{inverseMean} in the general form of \Eq{general}, we obtain
\begin{equation*}
  \GD f=\lrb{\frac{1}{\bq'\cdot\bz}-\frac{1}{\bq\cdot\bz}}
  		+\lrb{\frac{1}{\bq'\cdot\bz'}-\frac{1}{\bq'\cdot\bz}},
\end{equation*}
from which we can derive an expression that contains the standard partial change terms of the classic Price equation, normalized by the average values of $z$, as
\begin{equation}\label{eq:generalD}
  \GD f=-\frac{1}{\bq'\cdot\bz}\lrb{\frac{\bdq\cdot\bz}{\zbar}+\frac{\bq'\cdot\bdz}{\zbar'}}.
\end{equation}
Using the partial change notion in \Eq{partialz}, we can write
\begin{equation*}
  \GD f=-\frac{1}{\bq'\cdot\bz}\lrb{\frac{\GD_q\zbar}{\zbar}+\frac{\GD_z\zbar}{\zbar'}}.
\end{equation*}
This simple form emphasizes the partial change expression. In many applications, these partial change terms associate with the direct effects of selection on frequency changes and the indirect effects of changes in measurement values during transmission.

\section{Recursive Multilevel Expansion}

The classic Price equation can be expanded recursively \autocite{hamilton75innate}. In \Eq{canonical}, the right-hand $\bdz$ term is the vector $\GD z_i$ over the index, $i$. Each $\GD z_i$ term may itself be considered as the change in the average value of $z$ within the $i$th subset or group. Each of those terms may be analyzed by the equation, in which
\begin{equation*}
  \GD\zbar_i=\bdq_i\cdot\bz_i+\bq_i'\cdot\bdz_i.
\end{equation*}
Each group indexed by $i$ contains the next level of nested subsets. For example, $\bz_i=z_{1|i},z_{2|i},\dots,z_{1|m_i}$. Multilevel expansions have provided insight into many applications \autocite{hamilton75innate,frank98foundations,hilbert13linking}.

We can hierarchically expand the generalized form in \Eq{general}. In the third right-hand side term of that equation, we write
\begin{equation*}
  f(\bq',\bz')=f(\bq',\bz+\bdz),
\end{equation*}
in which $\bdz=\bz'-\bz$. Now we have a $\bdz$ term, to which we can apply the classic Price equation.

In the example of \Eq{generalD}, the $\bdz$ term appears in the numerator of the final right-hand side term. That term can be expanded to evaluate hierarchically nested levels by the methods outlined here.

\section{Application}

Several widely used population statistics do not match the standard expected value form of the classic Price equation \autocite{bullen03handbook}. Our generalization provides a way to partition changes in such statistics into frequency changes and measure changes. We discuss the harmonic mean and the numbers equivalent of a diversity index.

The harmonic mean of a measure $y_i$ has the form of \Eq{inverseMean} when we define $z_i$ as $1/y_i$. The expression in \Eq{generalD} partitions changes in the harmonic mean into partial frequency change and partial measure change components. 

Harmonic means arise is various population biology applications. The carrying capacity of a polymorphic population can often be approximated by a weighted harmonic mean of the carrying capacity of each type \autocite{anderson71genetic}. Optimal residence times of consumers may vary with the harmonic mean of the quality of resources in a given patch \autocite{calcagno14the-functional}. When mating occurs in local patches, the inbreeding level often follows the harmonic mean number of females per patch \autocite{herre85sex-ratio}.

The numbers equivalent diversity index can be expressed in the functional form of $f\lr{\bq,\bz}$ by defining $z_i=q_i^{p-1}$ and then writing
\begin{equation*}
  f\lr{\bq,\bz}=\lr{\bq\cdot\bz}^{\frac{1}{1-p}}.
\end{equation*}
This numbers equivalent form provides a general diversity expression that allows comparison of different diversity indices when expressed in common units \autocite{macarthur65patterns,hill73diversity,jost06entropy,jost07partitioning}.

Godsoe\autocite{godsoe19transmission} used the classic Price equation to analyze the simpler arithmetic mean diversity index $f=\bq\cdot\bz$ for $z_i=q_i^{p-1}$ for $p\ne1$. They demonstrated the value of partitioning changes in diversity into frequency change and measure change components to evaluate the role of selection in the dynamics of diversity. However, they could not generalize their results to the broader numbers equivalent form of diversity because the classic Price equation is limited to expected value statistics. We can now use the generalized Price equation to partition the change in the numbers equivalent diversity index.

\section{Conclusion}

The classic Price equation has been applied to a wide variety of problems \autocite{lehtonen20fifty}. Our results generalize the equation to population statistics that can be expressed as a function of frequencies and measurements. That generalization opens new problems for study.

\section*{Acknowledgments}

\noindent The project was supported by the Donald Bren Foundation (SAF) and the New Zealand Tertiary Education Commission CoRE grant to the Bio-Protection Research Centre (WG).

\bibliography{main}


\end{document}